\documentclass[12pt]{iopart}

\usepackage{epubtk}    
\usepackage{graphicx}  

\begin{document}

\title{Instrumentation for the Citizen CATE Experiment:
Faroe Islands and Indonesia}

\author{M J Penn$^1$,
R Baer$^2$,
R Bosh$^3$,
D Garrison$^4$,
R Gelderman$^3$,
H Hare$^3$,
F Isberner$^2$,
L Jensen$^5$,
S Kovac$^2$,
M McKay$^6$,
A Mitchell$^1$,
M Pierce$^5$,
A Ursache$^4$,
J Varsik$^7$,
D Walter$^8$,
Z Watson$^1$,
D Young$^9$,
and the Citizen CATE Team
}

\address{$^1$ NSO, 950 N Cherry Ave, Tucson AZ 85718}
\address{$^2$ Department of Physics, SIUC, Carbondale IL}
\address{$^3$ Department of Physics and Astronomy, WKU, Bowling Green KY}
\address{$^4$ Mathworks Inc, Natick MA}
\address{$^5$ Department of Physics and Astronomy, UW, Laramie WY}
\address{$^6$ Space Telescope Science Institute, Baltimore MD}
\address{$^7$ Big Bear Solar Observatory, Big Bear Lake, CA}
\address{$^8$ Department of Biological and Physical Sciences, SCSU, Orangeburg SC}
\address{$^9$ St Louis, MO}

\ead{mpenn@nso.edu}
\vspace{10pt}


\begin{abstract}
The inner regions of the solar corona from
1-2.5 Rsun are poorly sampled both from the ground
and space telescopes.
A solar eclipse reduces the sky scattered background intensity
by a factor of about 10,000 and opens a window to view this
region directly.
The goal of the Citizen
{\it Continental-America Telescopic Eclipse}
(CATE)
Experiment
is to take a 90-minute time sequence of calibrated white light images
of this coronal region using 60 identical telescopes spread
from Oregon to South Carolina during the 21 August 2017
total solar eclipse.
Observations that can address questions of coronal dynamics in
this region can be collected with rather modest telescope equipment,
but the large dynamic range of the coronal brightness requires careful
camera control.
The instruments used for test runs on the Faroe Islands in 2015
and at five sites in Indonesia in 2016 are described.
Intensity calibration of the coronal images is done
and compared with previous eclipse measurements
from November \& Koutchmy (1996)
and Bazin et al. (2015).
The change of coronal brightness with distance from
the Sun seen in the 2016 eclipse
agrees with observations from the 1991 eclipse
but differ substantially from the 2010 eclipse.
The 2015 observations agree with 2016 and 1991
solar radii near the Sun,
but are fainter at larger distances.
Problems encountered during these test runs are 
discussed as well the solutions which will be
implemented for the 2017 eclipse experiment.
\end{abstract}

\section{Introduction}

The path of totality of a solar eclipse will cross over
10 million homes in the USA during the late morning and early afternoon
on Monday 21 August 2017.
Tens of millions more people will travel to view the total
eclipse and hundreds of millions
more will directly view the partial eclipse.
Using broadcasts hundreds of millions of people will watch the
total eclipse, from school children through senior citizens. 
From one location during the 2017 eclipse the corona will
only be revealed for about 2.5 minutes;
this short time does not allow
detailed study of slower changes in the corona
(Lites et al., 1999).
From the moment the lunar shadow touches Oregon until
it leaves South Carolina 90 minutes of time will elapse.
The Citizen 
{\it Continental-America Telescopic Eclipse} (CATE) Experiment
will use 60 identical telescopes positioned across
the country to image the solar corona.
The CATE goal is to collect
calibrated white-light image of the solar corona
from about
$1~R_{sun}$
to
$2~R_{sun}$
with about 2 arcsecond pixels
every 10 seconds
continuously for 90 minutes.

Using data from the Spartan 201-01 mission in 1993,
Fisher and Guhathakurta (1995)
measured white light
polar plumes above the northern and southern solar coronal hole. 
These plumes extended from the lower limit of the occulting disk at
$R=1.25~R_{sun}$
up to over
$5~R_{sun}$
Simultaneous ground-based measurements from Mauna Loa suggested
that the plumes extended down to
$R=1.16~R_{sun}$.
The directions of the plumes,
while appearing roughly radial,
did not intersect
the center of the solar disk
but rather seemed to originate at higher latitudes. 
Later work by
DeForest and Gurman (1998)
traced these structures
down to magnetic features at the solar poles using SOHO EIT 171A data,
and measured a size of between 3-5 arcsec. 
The CATE data will measure these structures in white light
with 2 arcsec pixels to very
low heights of
$R=1.05~R_{sun}$,
and out to the edge of the field-of-view at
$2~R_{sun}$.
Using simultaneous magnetograms from other telescopes,
these polar plume structures can be
traced using the continuum signal from coronal
electron density enhancements back to the Sun with better resolution
than previous studies.

At solar minimum structures called polar plumes are clearly
visible above the magnetic north and south poles of the Sun.
These regions have been found to be very dynamic.
DeForest and Gurman (1998)
used SOHO EIT 171A observations to
observe outwardly moving density
enhancements traveling at
75-150 km/s velocity,
showing brightness changes of
5 to 10\%,
and displaying periodicity at
10-15 minute periods.
Cranmer (2004)
estimated
3 to 15\%
variations in the electron density in these events.  
Using UVCS observations
at two alternating heights in the corona 
$R=1.9~R_{sun}$
and
$2.1~R_{sun}$),
Ofman et al (2000)
found quasi-periodic variations of
5 to 10\%
in polarized brightness traveling radially at
210 km/s
with periods between
6.5 and 10.5 minutes. 
Morgan et al (2004)
used Lyman alpha data out to 
$2.2~R_{sun}$
to find oscillations with
7 to 8 minute periods.
With SUMER disk observations of Ne VIII emission,
Gupta et al (2012) 
found
5-10\%
intensity oscillations traveling at
60 km/s
with a period of
14.5 minutes. 
From eclipse observations,
Pasachoff (2009)
found changes in the corona above the southern solar pole.
But since this observation used only two images taken 19 minutes apart,
it is difficult to find systematic radial motion.
With 540 images taken at 10 second cadence across 90 minutes,
the CATE data will have direct applications to the study of polar 
plume dynamics.
Periodicities at the 15 minute time scale will be fully sampled,
and the velocities and accelerations of these events
will be measured from
$R=1.05~R_{sun}$
out to at least
$2~R_{sun}$.
The CATE data will be sensitive to transverse velocities
of roughly
0.8 to 145 km/sec
(3pix/90 min to 1 pix/10 sec)
and will easily measure these events.

In addition to showing polar plumes, the solar minimum corona
which will be seen during the 2017 total solar eclipse will likely show
several prominences.
The interaction of the hot corona with the cold prominence plasma 
has been studied recently.
With the Hinode SOT instrument
Berger et al (2007)
show upwardly moving hot gas parcels,
thought to be Rayleigh-Taylor instabilities in the prominences. 
Typical sizes which were observed were about
2250 km,
and these features showed upward speeds of roughly
20 km/s. 
Using lower resolution white-light images of the hot coronal plasma,
Druckmuller et al (2014)
examined a static structure observed near a prominence called a smoke ring.
The authors speculate that these structures are related
to the RT instabilities seen in prominences. 
The CATE data will reveal and motions of these new coronal structures
during the 90 minutes of the eclipse,
and will have a transverse velocity
sensitivity that covers the expected
20km/s motion. 
The CATE observations may reveal how
the instabilities seen in prominences interact with the hot corona
and produce density enhancements or depletions such as these smoke rings.

In preparation for the 2017 CATE experiment, two test runs have been
completed at the total solar eclipses in 2015 and 2016.
Equipment similar to the 2017 instrument was used in each
case by citizen scientists and first-time eclipse observers.
We describe the results of these tests and plans for the 2017 
experiment in the following paper.

\section{CATE Instruments}

The overall goal of the project is to develop a low-cost instrument
from off-the-shelf components that are readily available to citizen scientists,
while maintaining the scientific capability required to 
address the key science questions discussed previously.

\subsection{General}

A detailed description of how the science goals for the CATE data map
into instrument requirements is given in
Penn, Baer \& Isberner (2015),
but we review the main constraints here. 
For a simple and inexpensive design with high throughput
that will remain flexible for use after the eclipse,
the instrument camera will be at prime focus. 
In order to resolve 2 arcsec structures in the corona
at a wavelength of 600nm, the aperture of the telescope
must be greater than 60mm,
but reasonable cost constraints and portability requirements
limit the aperture to less than 120mm. 
For an image across a large field-of-view
of roughly 4000 x 4000 arcsec
that has good chromatic correction,
the objective lens
should be about f/5 or slower. 
Finally by examining currently available detectors
in this price range,
the objective focal length should be
shorter than about 800mm in order to fit the desired FOV on the detector. 

\subsection{Telescopes, Mounts and Cameras for the 2015 and 2016 instruments}

In both 2015 and 2016 eclipse experiments, a German Equatorial
mount from Celestron, Inc. (model CG4) and a battery powered RA drive
was used to track the Sun during the eclipse. 
This was the only item that was the same in these two experiments.
The mounting legs were modified for
ease of transport during the 2016 eclipse.

In the 2015 eclipse, the telescope used was a Lunt 80mm
diameter 560mm focal length refractor (model LE-80 OTA)
with an ED Doublet lens. 
The camera was the Point Grey Grasshopper 3 USB3 CMOS device
using a
2048x2048 CMOSIS CMV4000 array (model GS3-U3-41C6M-C)
with
5.5 micron pixels. 
The predicted image scale is
2.03 arcseconds per pixel.
The control computer had a
1.9GHz processor running Windows~8,
with
4Gbyte RAM
and a
1 TByte disk drive.

For the 2016 eclipse, the telescope used was a Daystar 80mm
diameter 480mm focal length refractor (model 480E)
During testing of the telescope we noticed residual chromatic aberration
and added a
Wratten \#58 filter
(Peel, 2009)
in the optical beam just
in front of the camera
to sharpen the image by minimizing the focus variation over 
the restricted bandpass. 
The camera was the
Point Grey Grasshopper 3
USB3 CMOS device
using a
2448x2048 Sony IMX250 array (model GS3-U3-51S5M-C)
with 3.45 micron pixels.
The predicted image scale is 1.48 arcseconds per pixel.
The control computer had a 1.7GHz dual processor running Windows~10,
with 16~Gbyte RAM and a 105~Gbyte solid state drive.

\subsection{Data Collection software: exposure sequence, dead-time, duty cycle}

The data collection software varied dramatically between
the tests at the two eclipses. 
In the 2015 eclipse, the demonstration software Flycapture
provided by Point Grey was used to collect calibration and coronal images.
The eclipse observing plan was 
to expose the camera for a short exposure value
for the first 30 seconds of totality to capture the inner corona,
then to manually switch the exposure time to a longer value
and observe the outer corona for another 30 seconds. 
Unfortunately, clouds prevented any images from being collected
with the longer exposure value. 

For the 2016 eclipse calibration data was collected using a
freeware routine called Firecapture.
Dark and flat field images (of the daytime sky) were collected. 
A drift scan was run where the RA drive motor was turned off
and images were collected once per second as the solar image drifted
across the FOV; in this way it is possible to determine
the direction of geocentric West in pixel coordinates for each site.
The control computer clock was set to GPS time within 30 minutes of 
the start of totality at each location;
tests showed that the CPU clocks drifted about 50 milliseconds
per hour,
so that the times for the exposures at each site would
be calibrated to about 25 milliseconds.
During totality, a data collection program written in
Matlab used the hardware trigger mode of the 51S5M-C to take
a set of 7 different exposures. 
Exposure times of
0.4, 1.3, 4.0, 13, 40, 130 and 400 milliseconds
were used to capture the expected dynamic range of the corona
across the FOV. 
The camera expose pulse was generated by an Arduino UNO R3
microcontroller
providing an output strobe
into the GPIO input of the 51S5M-C camera.
The Arduino sketch produced pulses for
each of the 7 exposures, with 
a 30 millisecond delay and between each exposure and a
a longer 200 millisecond delay between each cycle.
The longer cycle delay time was required in order for
the control computer to write all seven images to the SSD. 
Overall during 968.7 milliseconds of time the camera
was exposing for 588.7 milliseconds,
representing a 61\% duty cycle.
Throughput to the computer disk was measured at about 7 frames per second.
The image display during totality lagged the data collection,
and so attempts to update the telescope pointing at the
HM Volendam site (to correct for ship motion)
were impossible.

\section{2015 Faroe Islands}

The March 2015 eclipse observations were suggested by Fred Isberner
in August of 2014.
Isberner trained with the eclipse instrument for just a few days
in February of 2015,
and then he traveled to the eclipse at his own expense during a vacation.

\subsection{Site, conditions, instrument}

The path of totality for the total eclipse of 20 March 2015
crossed the northern Atlantic Ocean,
intersection land at only the Faroe Islands and Svalbaard. 
Observations with the first CATE prototype telescope
were taken at the Vagar Airport Hotel in Sorvagur
on the island of Vagar in the
Faroe Islands at 62.0669N, and 7.2812W. 
The observers were Fred Isberner and Howard Harper. 
Conditions at this location were very poor,
with rain and clouds during most of the partial phase and
only a few short partly cloudy gaps before and during totality. 
Duration of totality at this location was was predicted to be 137~seconds.

Through the partly cloudy conditions a sequence of over 800 
images of the partially eclipsed Sun were taken with a 
neutral density solar filter covering the primary lens
between second and third contact.
The extinction factor of this filter was estimated by comparing 
the observed solar intensity with the filter on to the
observed intensity of the Moon, Mars, Jupiter and Saturn 
with the filter off on 30 January 2015.
A linear fit of the observed intensity ratios to the predicted
intensity ratios suggests an extinction factor of
about
$1.05 \times 10^{5}$.
These images are used to derive the measured solar disk brightness
for calibrating the coronal images.
The solar limb was fit and the radius was measured at
477.02 pixels
which gives an image scale of 2.02 arcseconds per pixel.
After second contact, the same exposure value was used to capture
37 useful images of the inner corona after removing the neutral density filter.
The observing plan was to take a second image sequence during totality with 
a longer exposure time, but the weather conditions prevented this sequence.
The 37 short exposure images were co-aligned and then coadded.
The lunar limb was fit and the radius was measured to be
497.50 pixels
which again suggests an image scale of 2.02 arcseconds per pixel.
Using the center of the moon in this summed totality image
as the origin,
a normalized radial-graded filter 
(NRGF, Morgan, Habbal \& Woo, 2006)
was computed to determine the average intensity with radius,
and to filter the eclipse image to enhance the coronal structure.
The filtered image is shown in Figure~1.

\epubtkImage{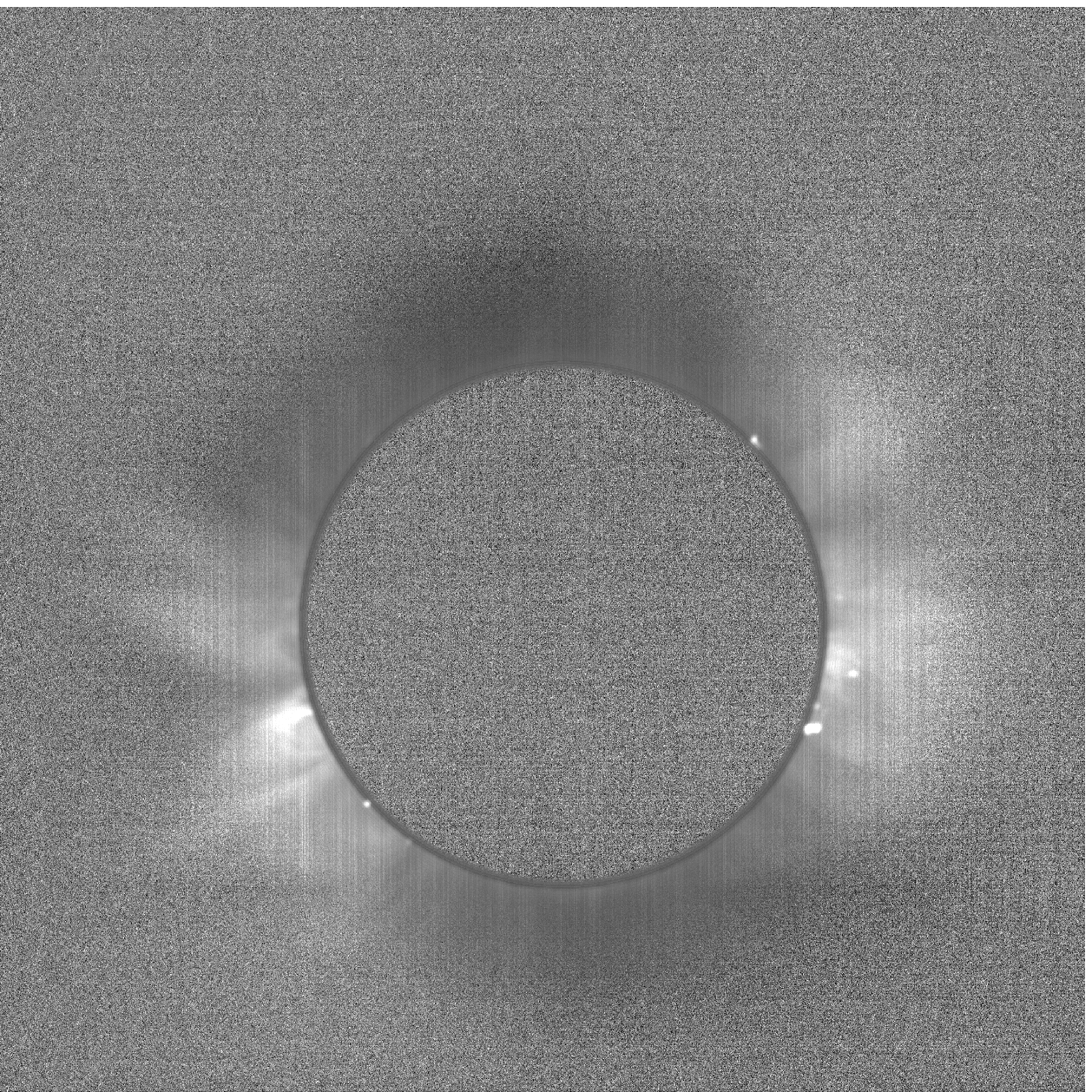}{%
\begin{figure}[htb]
  \centerline{\includegraphics[width=1.0\textwidth]{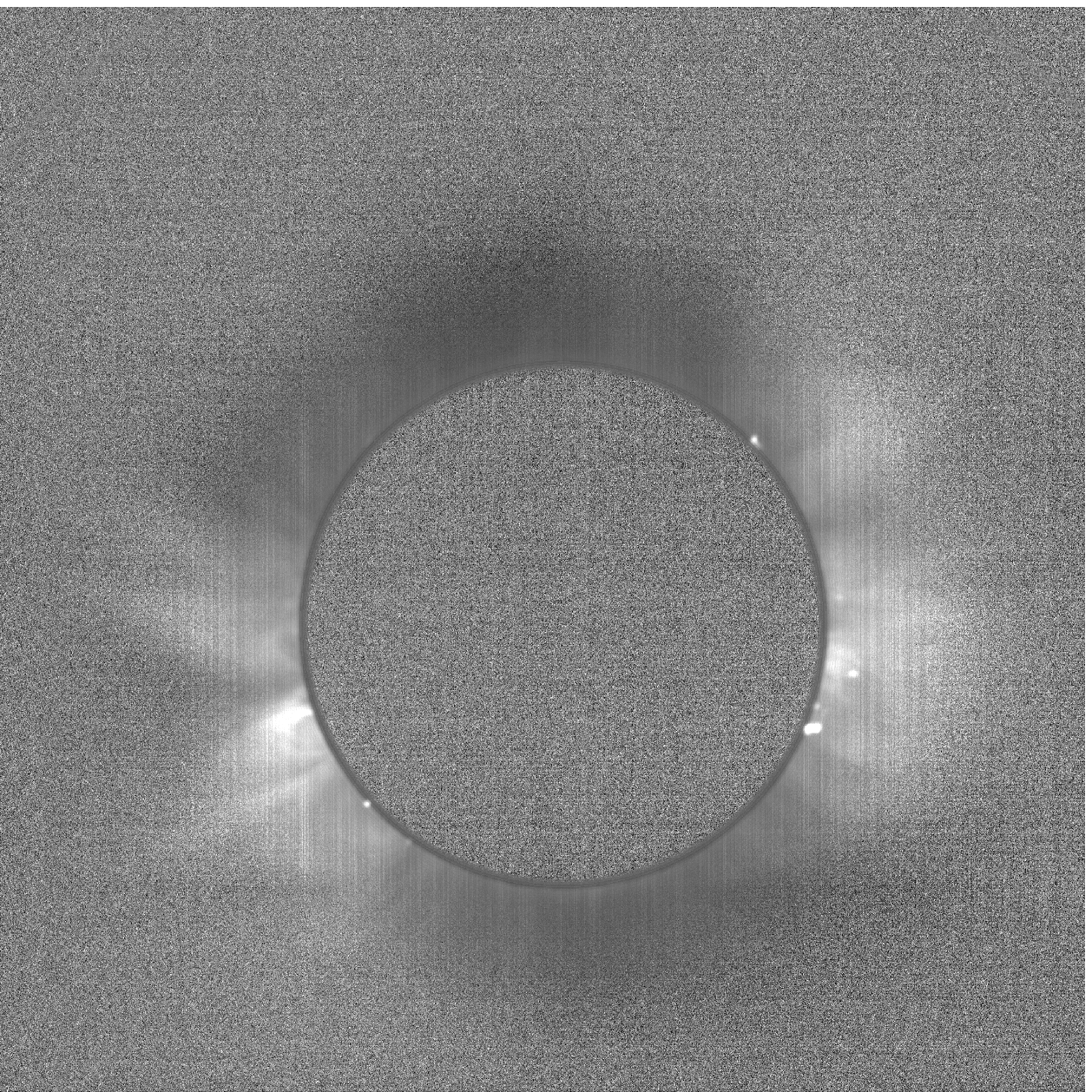}}
  \caption{Image of the solar corona from Vagar on the Faroe Islands
during the 2015 eclipse taken by Fred Isberner.
A normalized radial-graded filter has been applied to the sum of
31 co-aligned images from the observations.
}
  \label{fig:figure01}
\end{figure}}

\section{2016 Indonesia Network}

With support from NASA funding,
four university students were selected and traveled to the
National Solar Observatory in Tucson in January 2016.
For two days the students assembled their telescopes,
took sample solar data with the cameras and laptops,
and learned some basic ideas about the solar corona and eclipses.
None of the students had observed a solar eclipse previously,
and none had traveled outside of the USA.
A fifth group of citizen scientists traveling to the eclipse
at their own expense volunteered to take data as well.

\subsection{Sites, conditions, instruments}

The Indonesian eclipse experiment consisted of
five sites along the path of the total eclipse. 

The first site was in Tanjung Pandan on the island of Belitung.
Data collection occurred at
2.7433S, 107.6233E
in a park along the west coast of the island bordering the Java Sea.
The observers at this site were Sarah Kovac and Bob Baer.
Observing conditions were clear with small clouds.
There were no instrument problems and the images were focused well.  
Duration of totality was predicted to be 124~seconds.

The second site was
located in the city of Tanah Grogot
atop the RSU P.Sebaya hospital at
1.8729S, 116.1790E.
The observers at the site were Dr. Michael Pierce and Logan Jensen.
The weather conditions were calm but with obstructing clouds during totality.
Instrument behavior during the parts of the procedure performed was good,
however no data was acquired due to obscuring rain clouds. 
Duration of totality was 153~seconds.

The third site was located on the cruise ship MS Volendam
operated by Holland America Line. During totality the ship
was located roughly at
1.5983S, 118.0433E
in the Makassar Strait between Kalimantan and Sulawesi. 
The observers at this site were John Varsik, Fred Isberner and David Young.
Observing conditions were clear. 
Instrumentation problems during setup resulted in defocussed images,
saturated calibration data,
and
incorrect time-of-day values for images since the GPS software failed.
Finally waves during totality resulted in image motion of up
to several degrees during the observations.
Duration of totality was 160~seconds.

The fourth site was on the island of Sulawasi in Palu.
Observing occurred on the roof of a building on the eastern outskirts
of the city at approximately
0.8883S, 119.9100E
The observers at this site were Richard Gelderman and Honor Hare.
Observing conditions were clear with rare clouds.
The calibration images were over exposed before the eclipse began.
Also, the eclipse images were highly defocused and the sun off center. 
Duration of totality was 119~seconds.

The fifth site was island of Ternate.
Data collection occurred
at roughly
0.7958N, 127.3614E 
on a hotel room balcony on the east coast of the island.
The observers of this site were Myles McKay and Don Walter.
Observing conditions were cloudy but had a small break in
the clouds during totality.
There were no instrument problems,
but the images are defocused
since very little time was availabel to focus the telescope
during the partial phase of the eclipse due to the clouds.
Duration of totality was predicted to be 160~seconds.


During the partly cloudy conditions at Tanjung Pandan, calibration data
was obtained between first and second contact using a neutral
density filter.
The solar limb was fit and measured an image scale of
1.48 arcseconds per pixel.
The center was saturated.
Data from the HM Volendam had no clouds during the calibration or 
totality observing phases.
The solar limb was fit and measured an image scale of
1.49 arcseconds per pixel.
The totality data from Tanjung Pandan showed the best focus and image
stability.
Roughly 1038 coronal images were collected during totality.
A set of 7 exposures were used to make a high-dynamic range image
by subtracting dark exposures,
excluding saturated pixels and scaling by the correct exposure time.
In pixels where several images contained valid data, the scaled values
were averaged.
A fit to the lunar limb was done, resulting in
an image scale measurement of 1.49 arcseconds per pixel.
Figure~2 shows a sample of the data; it is a high-dynamic range image 
constructed from seven individual exposures with a NRGF filter applied 
to bring out coronal structures.

\epubtkImage{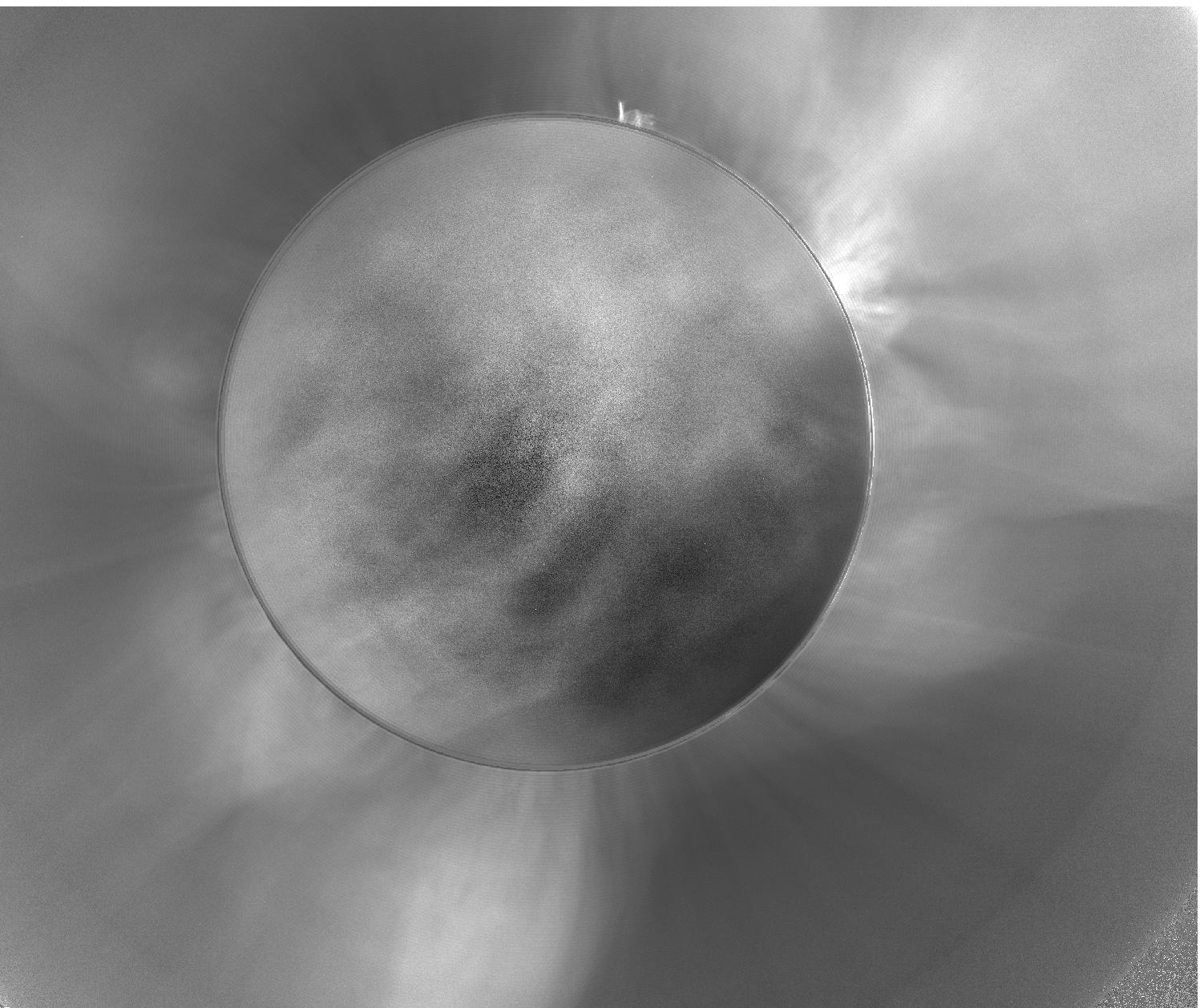}{%
\begin{figure}[htb]
  \centerline{\includegraphics[width=1.0\textwidth]{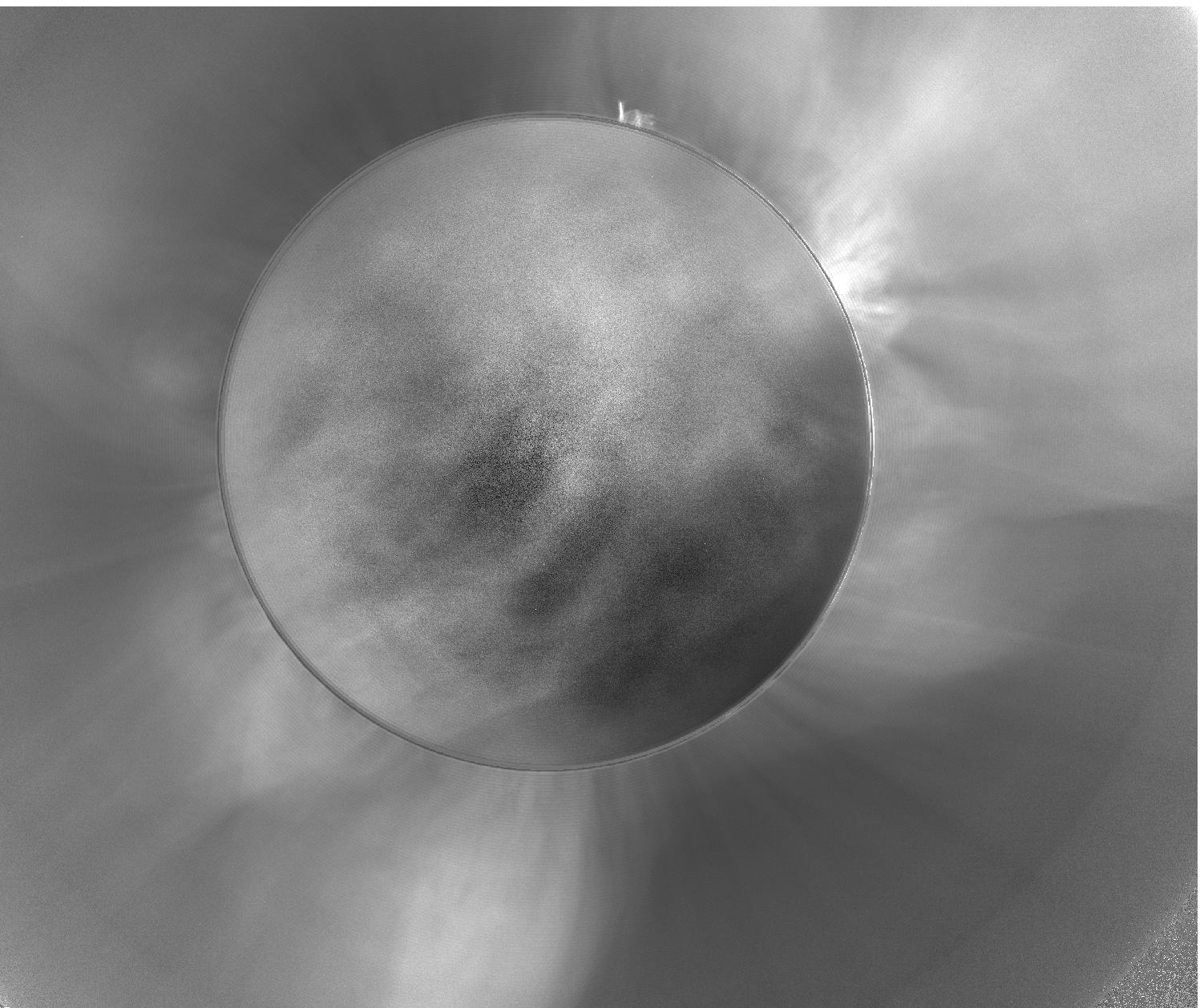}}
  \caption{Image of the solar corona from
the Indonesian 2016
eclipse taken by Bob Baer and Sarah Kovac.
A normalized radial-graded filter has been applied to the
high-dynamic range image composed from a single set of seven exposures.
}
  \label{fig:figure02}
\end{figure}}

\section{The Intensity Calibration}

With intensity images of the solar disk taken shortly before third contact,
and with a measurement of the extinction provided by
the neutral density filter,
calibrated measurement of the coronal brightness is possible.
This is the goal for the data collected at all the sites in the 2017 
eclipse experiment.

During tests in 2015 and 2016,
the extinction of the neutral density filters were measured using
night-time observations as discussed in the previous sections,
and images of the partially eclipsed solar disk were taken before totality.
Unfortunately several complications were encountered.
From the fives sites of coronal data,
three had partly cloudy conditions,
two had saturated pixel values at the center of the solar disk,
and one had images taken well after first contact which do not include
the center of the solar disk.
In this section we discuss methods to overcome some of these problems.

For the sites with partly cloudy condition
(Faroe Islands, Tanjung Pandan and Ternate)
the only possibility is to assume that the calibrated data
had equally cloudy conditions.
This assumption is probably not correct, but it is the best that can be done.

In order to find intensity values at disk center when the camera
pixels were saturated, 
and to estimate the sun-center brightness in late phases of the
partial eclipse,
we make the assumption that the measured limb darkening with each instrument
is identical. 
With the four sites from Indonesia this is a good assumption since they are
each measuring the solar disk at the same wavelengths, 
those transmitted by the
Wratten \#58 filter.
It is not strictly correct to directly compare the Faroe Islands
data with the Indonesian data since the wavelengths sampled are different
and the expected limb darkening function is not identical,
but we make the comparison anyway to estimate
the disk center brightness for
the Faroe Islands data.
The comparison of the measured solar disk intensity
for the Faroe Islands (Vagar), Volendam and Tanjung Pandan data is shown
in Figure~3.

\epubtkImage{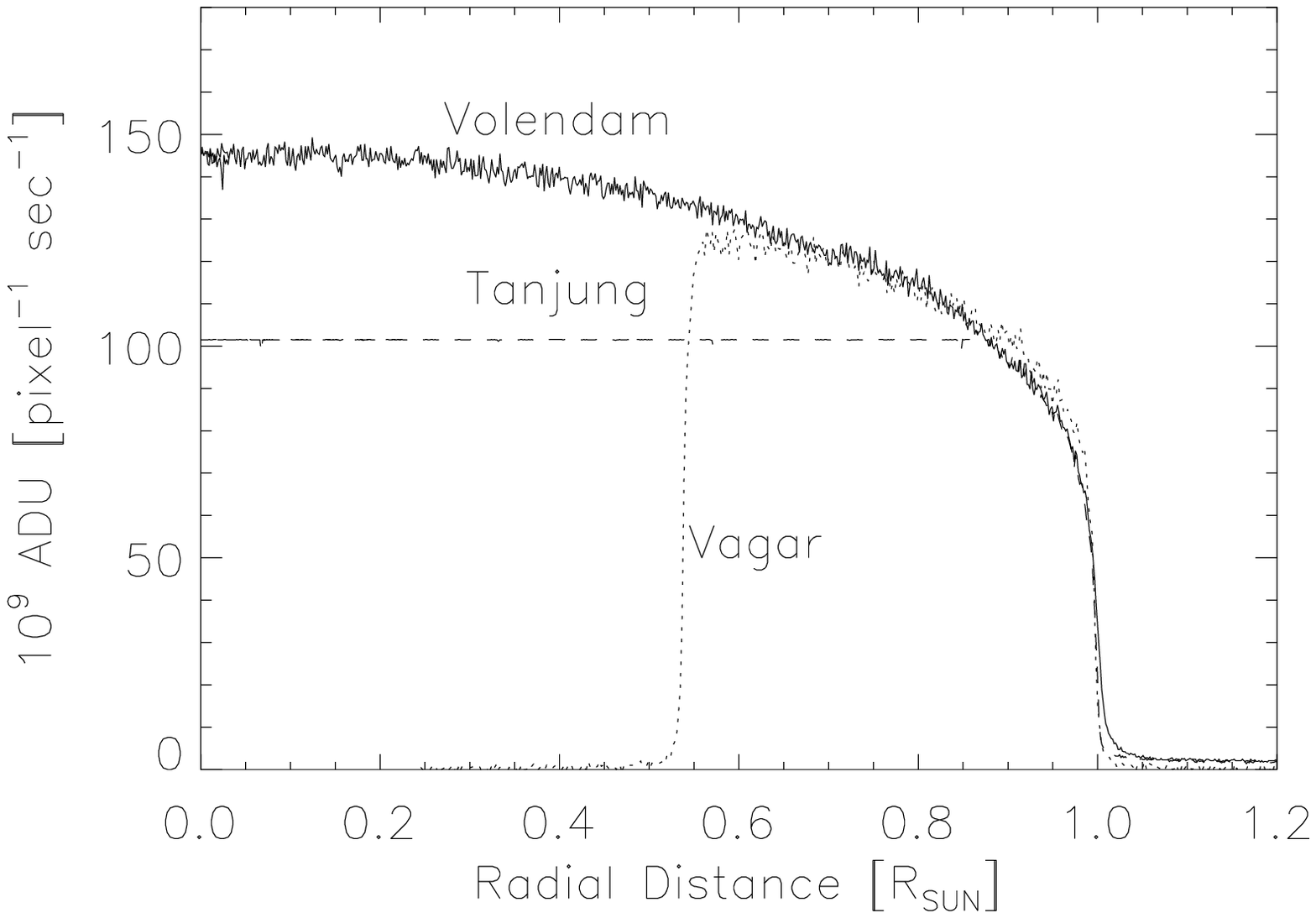}{%
\begin{figure}[htb]
  \centerline{\includegraphics[width=1.0\textwidth]{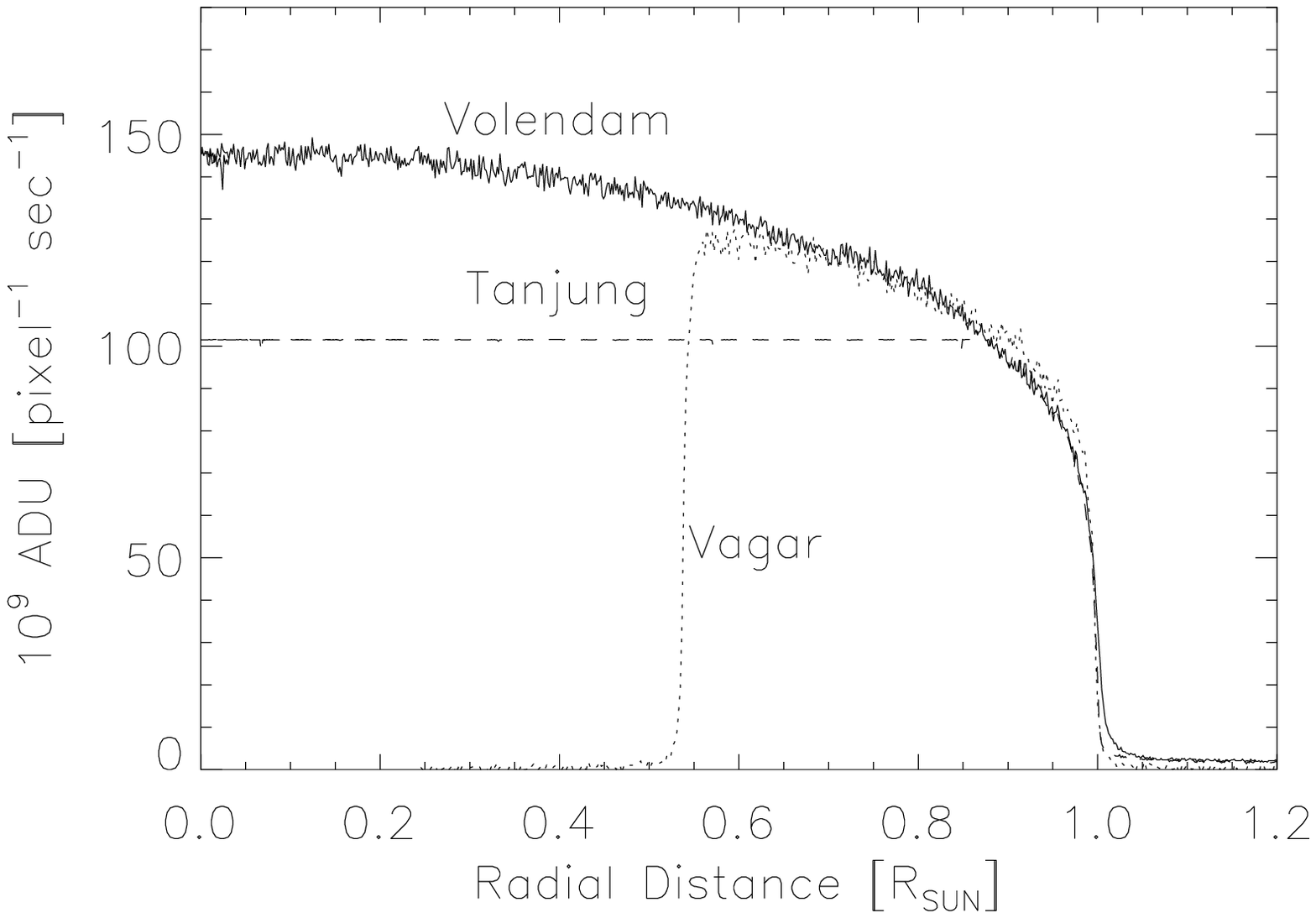}}
  \caption{Solar limb profiles from Volendam, Tanjung, Vagar.
The Volendam calibration image captures the solar disk from sun center to 
the limb and beyond;
the Vagar image was taken later in the eclipse and
the lunar limb blocks the Sun below about
$0.55~R_{sun}$
the Tanjung image saturates on the disk from the sun center out to about 
$0.87~R_{sun}$.
}
  \label{fig:figure03}
\end{figure}}

The Tanjung image was scaled to the Volendam image using intensites
from
$0.9~R_{sun}<r<0.97~R_{sun}$, 
and the Vagar image was scaled across the range of from
$0.6~R_{sun}<r<0.97~R_{sun}$. 
While the Volendam and Tanjung images were taken at the same wavelengths 
and have very similar limb darkening profiles,
the Vagar image was taken at a longer wavelength and only approximately
matches the Volendam limb darkening.
Using the measured value for the neutral density filter extinction of 27000,
accounting for a 38mm aperture stop used over the telescope primary lens,
and accounting for exposure time,
the Volendam calibration images suggest a count rate of 
145$\times 10^{9}$ ADU pixel$^{-1}$ second$^{-1}$.
Accounting for the measured scaling factor of 1.092 between the
Volendam and the Tanjung images at the solar limb,
the Tanjung count rate is extrapolated to be
158$\times 10^{9}$ ADU pixel$^{-1}$ second$^{-1}$
at solar disk center.
Finally, after fitting the Vagar data to Volendam, 
the extrapolated count rate at disk center for the 2015 data is
1.90$\times 10^{9}$ ADU pixel$^{-1}$ second$^{-1}$.

\section{Average Coronal Brightness}

Using the mean intensity of the corona computed during the NRGF image filtering,
and scaling by the intensity calibrations measured from the solar disk images,
we compute the calibrated azimuthally averaged coronal brightness.
The measured Vagar coronal brightness is about 
a factor of eight brighter than the Tanjung brightness;
this is likely to have been caused by a change in the camera
gain setting
during the totality observations 2015.
After accounting for this offset
the coronal brightness behaves identically in the two data sets from
about
$1.07~R_{sun}$
to
$1.8~R_{sun}$,
after which the 2015 coronal brightness drops faster with increasing distance.
Figure~4 plots these coronal brightnesses,
but each observation has been scaled differently;
the brightness from Vagar has been multiplied by a factor of 6.3
and the brightness from Tanjung has been multiplied by a factor of 79.
Also included on this plot are coronal intensity values measured
from the 1991 eclipse by 
November \& Koutchmy (1996),
and equatorial and polar coronal brightnesses reported by
Bazin et al. (2015).
The Tanjung data precisely fit the values from
November \& Koutchmy (1996)
when they are multiplied by a factor of 63,
but the Vagar values drop more quickly than any of the
previously reported measurements after a solar radial distance of
about
$1.8~R_{sun}$
perhaps due to the cloudy conditions.
It is likely that the differences in the absolute brightness calibrations
are caused by a combination of
incorrectly measuring the extinction of each 
neutral density filter
and by the clouds at both eclipse sites.

\epubtkImage{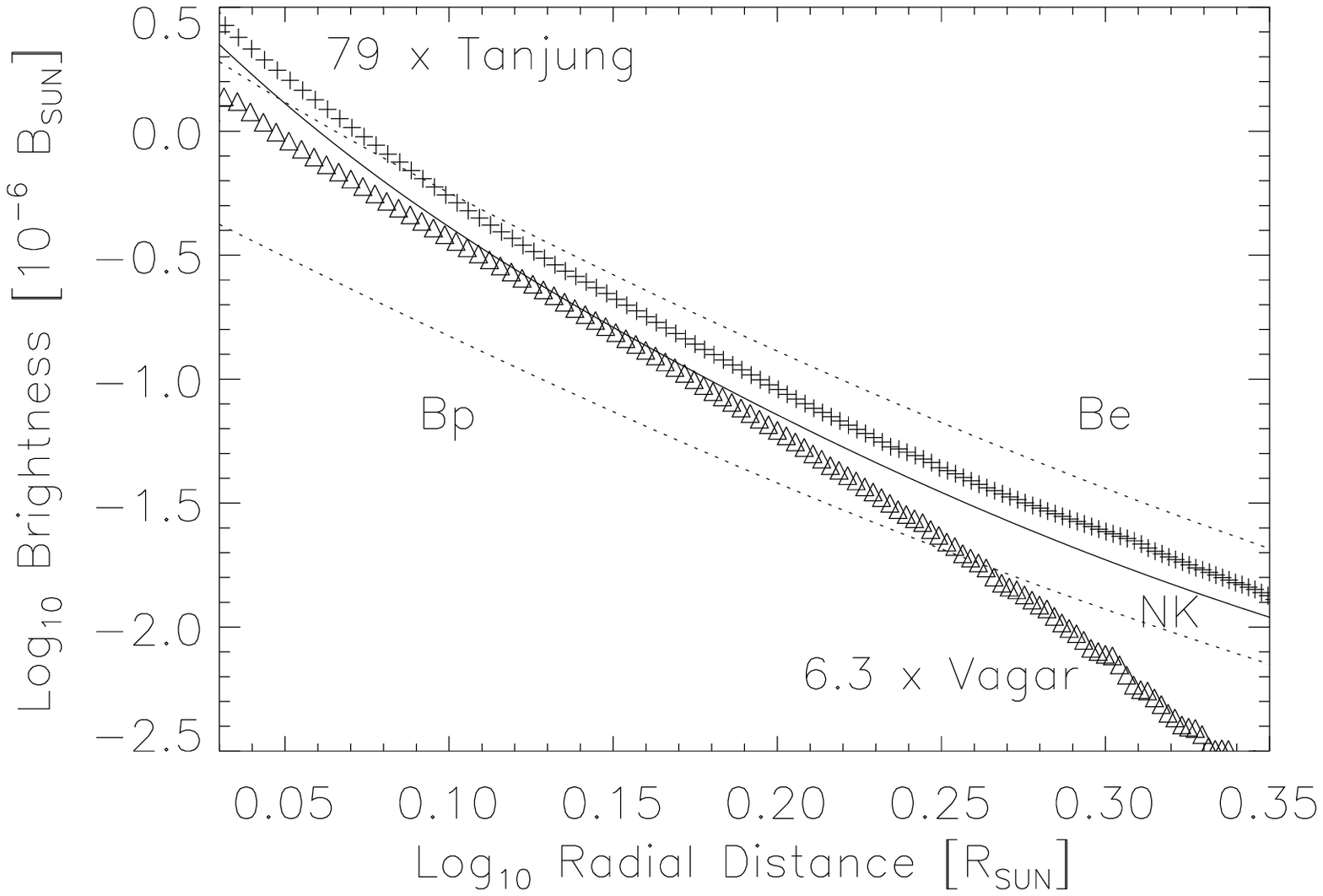}{%
\begin{figure}[htb]
  \centerline{\includegraphics[width=1.0\textwidth]{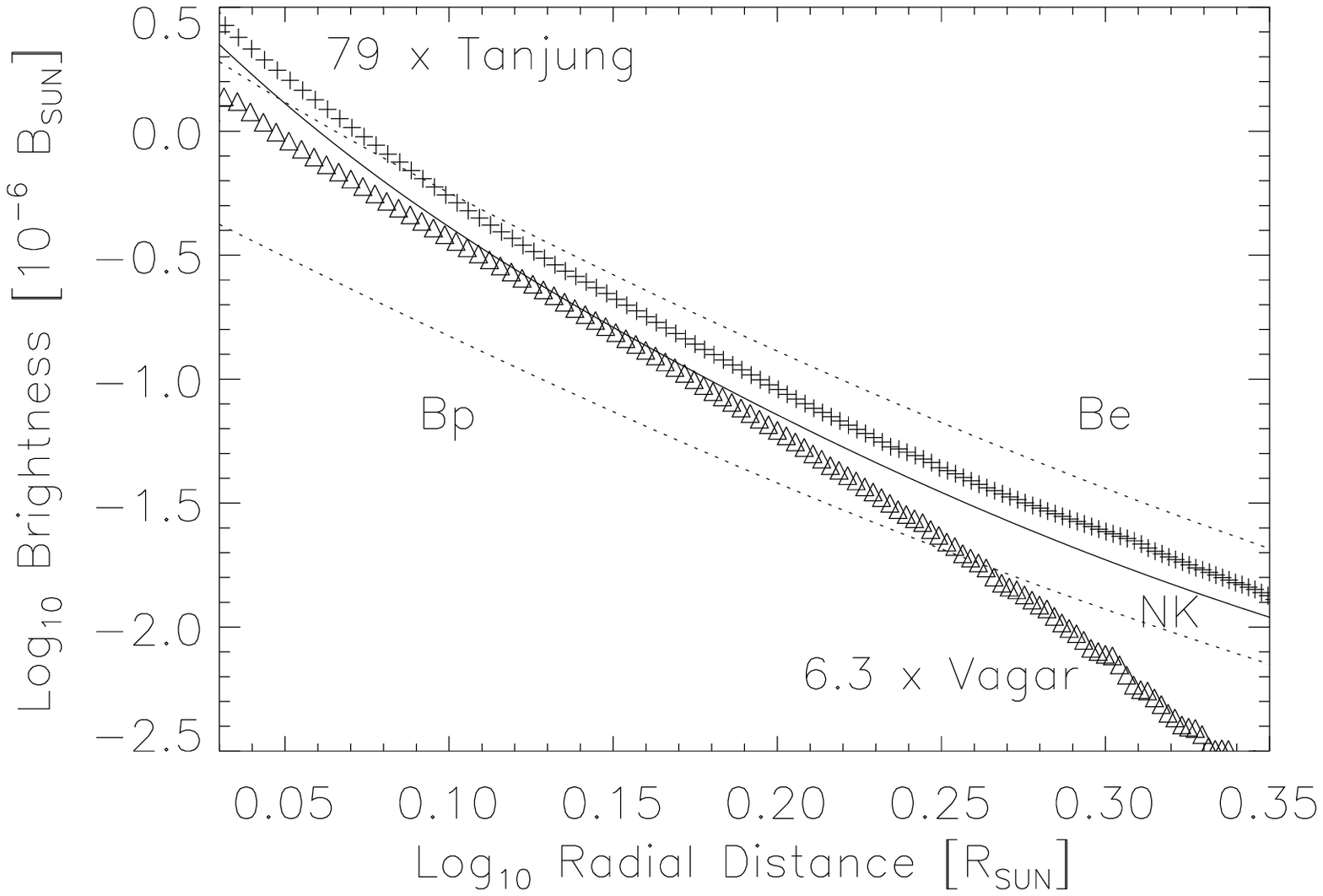}}
  \caption{The calibrated azimuthally averaged coronal brightness from the
2015 Vagar eclipse observations compared with the 2016 Tanjung observations.
The Tanjung data have been multiplied by a factor of 79,
and the Vagar data by a factor of 6.3.
Also shown are measurements from
November \& Koutchmy (1996) (solid line, NK)
and Bazin (2015) (Be is a reported 2010 equatorial coronal brightness,
and Bp is a reported 2010 polar coronal brightness).
The Tanjung data fit the 
November \& Koutchmy (1996)
measurements very closely if they are multiplied
by a factor of 63.
}
  \label{fig:figure03}
\end{figure}}

\section{The 2017 Eclipse Instrument}
Our experiences in 2015 and 2016 have pointed to a few problems 
which must be addressed with the 2017 instrument.
The telescope objective must have better chromatic correction,
the solar brightness data must be taken with
a calibrated filter and must not be saturated,
and the images of the Sun and the eclipse must be in focus.
Additionally, the operations of the
timing and GPS software should be more reliable.

\subsection{Instrument outline }

The overall instrument design will remain nearly identical to the 
2015 and 2016 telescopes; an 80mm diameter 400-500mm focal length
refractor will be used. 
Daystar Filters Inc has prepared a new glass prescription for the
objective lens which will result in much less chromatic aberration.
The manual focus mechanism will remain, as will the
German Equatorial tripod mount.
We expect to use the same (or similar) digital cameras and
very similar control computers.
\subsubsection{New Matlab GUI}
A new graphical user interface is being written in 
{\it Matlab}
which will aid the observers in collecting both the calibration
data and the totality data.
This GUI will address the three remaining problems uncovered during our
two years of tests.
First, an intensity histogram of the calibration images will be shown
to the observers,
and software checks will be introduced to make sure that the calibration 
images are not saturated.
The neutral density calibration solar filter will be supplemented with 
an aperture stop with a calibrated pinhole.
Second, a new focus routine has been written to give an objective feed-back
to the CATE observer.
The routine measures the intensity gradient across the solar image,
and provides a real-time measurement of the maximum value of the gradient.
The CATE observer will maximize this value during the partial phase of the
2017 eclipse in order to obtain a good focus for the telescope.
Finally a new shield board will be added to the Arduino camera controller.
This shield will have a GPS module which will obtain a GPS position and
time, but also provide a pulse-per-second output to the data collection
software.
In this way, each site in the 2017 network will collect an HDR image sequence
starting at even GPS seconds.
Thus all of the 2017 data will be temporally calibrated.

Software which will provide a quick-look at the 2017 eclipse data is
currently being developed.
The goal is to reduce the drift-scan data to determine the 
image rotation at each site,
and then to select a set of seven exposures from the totality
observations to construct an HDR image.
This HDR image will then be rotated and aligned at each
CATE site in 2017 using the control computer
(after the eclipse data has been backed-up)
and then uploaded to a web site to produce a movie of the
observations on the day of the eclipse.

\subsection{The 2017 CATE Sites}
The sites for the 2017 CATE experiment have been selected
and the volunteer observers for each location are established.
Sites west of the Mississippi River have good weather forecasts,
sites east of Mississippi have about a 50\% chance of clear skies.
Since the project would like to transfer ownership of the 
site equipment to the volunteers,
the CATE equipment is being funded with private and corporate
donations.
Currently about 20 sites are funded, and we are trying to
fund the equipment for all 60 sites soon.
The plan is to have the 2016 eclipse volunteers act as trainers
for the 2017 CATE volunteers.
The video and hands-on training program has been tested at a local
high school in Tucson AZ,
who will be sending a team of 8 students
to collaborate with another high school in Beatrice NE.
The volunteers for the 2017 eclipse range in age from 
middle-school students through retired professional solar astronomers.
The observing locations range from remote camp sites to college campus
locations where tens of thousands of visitors are expected.

\subsection{Data plan}
Each CATE site is expected to produce about 10 Gbytes of calibration
data and about 10 Gbytes of coronal images.
Many locations will not have any internet access,
and so it is expected that one copy of the data
will be mailed to NSO offices.
With a three-day delivery time, 
mailing the data results in about 
5 Mbyte per second transfer rate, which seems acceptable.
The data will be analyzed at the NSO centrally.

The quick-look images produced on each site laptop will be uploaded that
day to the NSO; each image will be less than 10 Mbytes.
A movie sequence of these images will be produced as soon as possible for 
education and public outreach purposes.

\section{Future use on night-time objects}
Since each of the CATE instruments will be transferred to the site
volunteers, it is important to have follow-up citizen science programs
for the telescopes so that the volunteers can continue to engage
in citizen science for astronomy.
A solar observing program is 
being investigated which would use an additional H-alpha filter,
tuned with the Arduino controller,
to produce full-disk dopplergrams for the chromosphere.
Observing cometary light curves provides valuable information about
evolution of comets but is infrequently supported at large facilities.
A follow-up project collecting photometry on comets is being prepared 
for the CATE telescopes.
Finally the AAVSO is developing an instrument package which is similar
in size and spatial resolution to the CATE instrument; 
a variable star observing program for the CATE instrument
is being developed as well.

\ack
The NSO is operated by AURA, Inc. for the NSF.
Student work was partially supported through the NSF REU program
with funding from AST-1460743.
This work was also supported by funding from NASA SMD grant NNX16AB92A.

\References

\item[] Bazin et al 2015 SF2A-2015, p259
\item[] Berger et al, 2008, ApJ 676, L89
\item[] Cranmer, 2004, ESA SP-547, 353
\item[] DeForest \& Gurman, 1998, ApJ 501, L217
\item[] Druckmuller et al., 2014, ApJ 785,14
\item[] Fisher \& Guhathakurta, 1995, ApJ 447, L139
\item[] Gupta et al. 2012, AA 546,A93
\item[] Lites et al. 1999, Solar Physics, 190, 185.
\item[] Morgan et al. 2004, ApJ 605,521
\item[] Morgan, H., Habbal, SR, Woo, R 2006, Solar Physics v236 p263-272
\item[] November, LJ \& Koutchmy, S 1996, ApJ 466 512
\item[] Ofman et al, 2000, ApJ, 529,592
\item[] Pasachoff et al 2009, ApJ 702,1297
\item[] Peel, A. 2009, CRC Handbook of Chemistry \& Physics, 90-th Edition, ed. D. R. Lide, CRC Press.
\item[] Penn, Baer \& Isberner, 2015, SAS, 34, p63.

\endrefs

\end{document}